\documentstyle[twoside,psfig]{article}

\catcode`\@=11
\long\def\@makefntext#1{
\protect\noindent \hbox to 3.2pt {\hskip-.9pt
$^{{\eightrm\@thefnmark}}$\hfil}#1\hfill}       

\def\thefootnote{\fnsymbol{footnote}}
\def\@makefnmark{\hbox to 0pt{$^{\@thefnmark}$\hss}}    

\def\ps@myheadings{\let\@mkboth\@gobbletwo
\def\@oddhead{\hbox{}
\rightmark\hfil\eightrm\thepage}
\def\@oddfoot{}\def\@evenhead{\eightrm\thepage\hfil
\leftmark\hbox{}}\def\@evenfoot{}
\def\sectionmark##1{}\def\subsectionmark##1{}}

\oddsidemargin=\evensidemargin
\renewcommand{\thefootnote}{\fnsymbol{footnote}}

\newcounter{sectionc}\newcounter{subsectionc}\newcounter{subsubsectionc}
\renewcommand{\section}[1] {\vspace{12pt}\addtocounter{sectionc}{1}
\setcounter{subsectionc}{0}\setcounter{subsubsectionc}{0}\noindent
    {\tenbf\thesectionc. #1}\par\vspace{5pt}}
\renewcommand{\subsection}[1] {\vspace{12pt}\addtocounter{subsectionc}{1}
    \setcounter{subsubsectionc}{0}\noindent
    {\bf\thesectionc.\thesubsectionc. {\kern1pt \bfit #1}}\par\vspace{5pt}}
\renewcommand{\subsubsection}[1] {\vspace{12pt}\addtocounter{subsubsectionc}{1}
    \noindent{\tenrm\thesectionc.\thesubsectionc.\thesubsubsectionc.
    {\kern1pt \tenit #1}}\par\vspace{5pt}}
\newcommand{\nonumsection}[1] {\vspace{12pt}\noindent{\tenbf #1}
    \par\vspace{5pt}}

\newcounter{appendixc}
\newcounter{subappendixc}[appendixc]
\newcounter{subsubappendixc}[subappendixc]
\renewcommand{\thesubappendixc}{\Alph{appendixc}.\arabic{subappendixc}}
\renewcommand{\thesubsubappendixc}
    {\Alph{appendixc}.\arabic{subappendixc}.\arabic{subsubappendixc}}

\renewcommand{\appendix}[1] {\vspace{12pt}
        \refstepcounter{appendixc}
        \setcounter{figure}{0}
        \setcounter{table}{0}
        \setcounter{lemma}{0}
        \setcounter{theorem}{0}
        \setcounter{corollary}{0}
        \setcounter{definition}{0}
        \setcounter{equation}{0}
        \renewcommand{\thefigure}{\Alph{appendixc}.\arabic{figure}}
        \renewcommand{\thetable}{\Alph{appendixc}.\arabic{table}}
        \renewcommand{\theappendixc}{\Alph{appendixc}}
        \renewcommand{\thelemma}{\Alph{appendixc}.\arabic{lemma}}
        \renewcommand{\thetheorem}{\Alph{appendixc}.\arabic{theorem}}
        \renewcommand{\thedefinition}{\Alph{appendixc}.\arabic{definition}}
        \renewcommand{\thecorollary}{\Alph{appendixc}.\arabic{corollary}}
        \noindent{\tenbf Appendix \theappendixc #1}\par\vspace{5pt}}
\newcommand{\subappendix}[1] {\vspace{12pt}
        \refstepcounter{subappendixc}
        \noindent{\bf Appendix \thesubappendixc. {\kern1pt \bfit #1}}
    \par\vspace{5pt}}
\newcommand{\subsubappendix}[1] {\vspace{12pt}
        \refstepcounter{subsubappendixc}
        \noindent{\rm Appendix \thesubsubappendixc. {\kern1pt \tenit #1}}
    \par\vspace{5pt}}

\newcommand{\textlineskip}{\baselineskip=13pt}
\newcommand{\smalllineskip}{\baselineskip=10pt}

\def\eightcirc{
\begin{picture}(0,0)
\put(4.4,1.8){\circle{6.5}}
\end{picture}}
\def\eightcopyright{\eightcirc\kern2.7pt\hbox{\eightrm c}}

\newcommand{\copyrightheading}[1]
    {\vspace*{-2.5cm}\smalllineskip{\flushleft
    {\footnotesize International Journal of Modern Physics C #1}\\
    {\footnotesize $\eightcopyright$\, World Scientific Publishing
     Company}\\
     }}

\newcommand{\publisher}[2]{{\begin{center}\footnotesize\smalllineskip
    Received #1\\
    Revised #2
    \end{center}
    }}

\def\abstracts#1#2#3{{
    \centering{\begin{minipage}{4.5in}\footnotesize\baselineskip=10pt
    \parindent=0pt #1\par
    \parindent=15pt #2\par
    \parindent=15pt #3
    \end{minipage}}\par}}

\def\keywords#1{{
    \centering{\begin{minipage}{4.5in}\footnotesize\baselineskip=10pt
    {\footnotesize\it Keywords}\/: #1
    \end{minipage}}\par}}

\newcommand{\bibit}{\nineit}
\newcommand{\bibbf}{\ninebf}
\renewenvironment{thebibliography}[1]
        {\frenchspacing
     \ninerm\baselineskip=11pt
         \begin{list}{\arabic{enumi}.}
        {\usecounter{enumi}\setlength{\parsep}{0pt}
     \setlength{\leftmargin 12.7pt}{\rightmargin 0pt} 
         \setlength{\itemsep}{0pt} \settowidth
    {\labelwidth}{#1.}\sloppy}}{\end{list}}

\newcounter{itemlistc}
\newcounter{romanlistc}
\newcounter{alphlistc}
\newcounter{arabiclistc}

\newcommand{\fcaption}[1]{
        \refstepcounter{figure}
        \setbox\@tempboxa = \hbox{\footnotesize Fig.~\thefigure. #1}
        \ifdim \wd\@tempboxa > 5in
           {\begin{center}
        \parbox{5in}{\footnotesize\smalllineskip Fig.~\thefigure. #1}
            \end{center}}
        \else
             {\begin{center}
             {\footnotesize Fig.~\thefigure. #1}
              \end{center}}
        \fi}

\newcommand{\tcaption}[1]{
        \refstepcounter{table}
        \setbox\@tempboxa = \hbox{\footnotesize Table~\thetable. #1}
        \ifdim \wd\@tempboxa > 5in
           {\begin{center}
        \parbox{5in}{\footnotesize\smalllineskip Table~\thetable. #1}
            \end{center}}
        \else
             {\begin{center}
             {\footnotesize Table~\thetable. #1}
              \end{center}}
        \fi}

\def\@citex[#1]#2{\if@filesw\immediate\write\@auxout
    {\string\citation{#2}}\fi
\def\@citea{}\@cite{\@for\@citeb:=#2\do
    {\@citea\def\@citea{,}\@ifundefined
    {b@\@citeb}{{\bf ?}\@warning
    {Citation `\@citeb' on page \thepage \space undefined}}
    {\csname b@\@citeb\endcsname}}}{#1}}

\newif\if@cghi
\def\cite{\@cghitrue\@ifnextchar [{\@tempswatrue
    \@citex}{\@tempswafalse\@citex[]}}
\def\citelow{\@cghifalse\@ifnextchar [{\@tempswatrue
    \@citex}{\@tempswafalse\@citex[]}}
\def\@cite#1#2{{$\null^{#1}$\if@tempswa\typeout
    {IJCGA warning: optional citation argument
    ignored: `#2'} \fi}}

\def\pmb#1{\setbox0=\hbox{#1}
    \kern-.025em\copy0\kern-\wd0
    \kern.05em\copy0\kern-\wd0
    \kern-.025em\raise.0433em\box0}

\def\fnt#1#2{\footnotetext{\kern-.3em
    {$^{\mbox{\scriptsize #1}}$}{#2}}}

\def\ps@myheadings{%
    \let\@oddfoot\@empty\let\@evenfoot\@empty
    \def\@evenhead{\slshape\leftmark\hfil}
    \def\@oddhead{\hfil{\slshape\rightmark}}
    \let\@mkboth\@gobbletwo
    \let\sectionmark\@gobble
    \let\subsectionmark\@gobble
    }
\font\tenrm=cmr10
\font\tenit=cmti10
\font\tenbf=cmbx10
\font\bfit=cmbxti10 at 10pt
\font\ninerm=cmr9
\font\nineit=cmti9
\font\ninebf=cmbx9
\font\eightrm=cmr8

\def\qed{\hbox{${\vcenter{\vbox{            
   \hrule height 0.4pt\hbox{\vrule width 0.4pt height 6pt
   \kern5pt\vrule width 0.4pt}\hrule height 0.4pt}}}$}}

\renewcommand{\thefootnote}{\fnsymbol{footnote}}    

\def\bsc{{\sc a\kern-6.4pt\sc a\kern-6.4pt\sc a}}   
\def\bflatex{\bf L\kern-.30em\raise.3ex\hbox{\bsc}\kern-.14em
T\kern-.1667em\lower.7ex\hbox{E}\kern-.125em X}

\begin{document}
\setlength{\textheight}{7.7truein}  

\thispagestyle{empty}

\markboth{\protect{\footnotesize\it Tu Yu-Song, A.O. Sousa, Kong Ling-Jiang
    and  Liu Mu-Ren}}
{\protect{\footnotesize\it Sznajd model with synchronous updating
on complex networks}}

\normalsize\textlineskip

\setcounter{page}{1}

\copyrightheading{}         

\vspace*{0.88truein}

\centerline{\bf Sznajd model with synchronous updating on complex
networks} \vspace*{0.035truein} \vspace*{0.37truein}
\centerline{\footnotesize Tu Yu-Song$^1$\footnotemark[1] , A.O. Sousa$^2$, 
Kong Ling-Jiang$^1$ and Liu Mu-Ren $^1$\footnotemark[2]} \baselineskip=12pt
\centerline{\footnotesize\it  $^1$ College of Physics and Information 
Engineering, Guangxi Normal University, } \baselineskip=10pt
\centerline{\footnotesize\it 541004 Guilin, China.}\centerline{\footnotesize\it \ e-mails: \footnotemark[1] 
tuyusong@sohu.com, \footnotemark[2] lmrlmr@mailbox.gxnu.edu.cn}

\vspace*{10pt}     
\centerline{\footnotesize\it $^2$ Institute for Computer 
Physics (ICP), University of Stuttgart,} \baselineskip=10pt
\centerline{\footnotesize\it Pfaffenwaldring 27, 70569 Stuttgart,
Germany.} \centerline{\footnotesize\it e-mail:
sousa@ica1.uni-stuttgart.de} \vspace*{0.225truein}
\publisher{(received date)}{(revised date)}

\vspace*{0.25truein} \abstracts{ We analyze the evolution of
Sznajd Model with synchronous updating in several complex
networks. Similar to the model on square lattice, we have found a 
transition between the state with no-consensus and the state with
complete consensus in several complex networks. Furthermore, by
adjusting the network parameters, we find that a large clustering
coefficient favors development of a consensus. In particular, in the 
limit of large system size with the initial concentration $p=0.5$ of 
opinion $+1$, a consensus seems to be never reached for the Watts-Strogatz 
small-world network, when we fix the connectivity $k$ and the rewiring
probability $p_s$; nor for the scale-free network, when we fix the
minimum node degree $m$ and the triad formation step probability 
$p_{t}$.}{}{}

\vspace*{5pt} \keywords{opinion dynamics, Sznajd model, small-world
  networks, synchronous updating, computer simulation.}


\vspace*{1pt}\textlineskip  
\section{Introduction }     
\vspace*{-0.5pt} \noindent Recently, Sznajd-Weron proposed a consensus
model\cite{1,2} (it's now called Sznajd model), which is a
successful Ising spin model describing a simple mechanism of
making up decisions in a closed community: A pair of nearest
neighbors convinces its neighbors of the pair opinion if and only
if both members of the pair have the same opinion; otherwise the
pair and its neighbors do not change opinion. The consensus model
of Sznajd has rapidly acquired importance in the new field of
computational socio-physics\cite{3,4}.

In Sznajd model there are two ways of updating the system states:
Asynchronous and Synchronous updating. While the asynchronous updating has 
been already analyzed considering the Sznajd model on an one-dimensional 
lattice, on a square lattice\cite{2}, on a triangular lattice\cite{5}, the 
dilute\cite{6}, on a three-dimensional cubic lattice \cite{7} and even 
on some networks\cite{7,8,9,10}, the synchronous updating has been only 
studied on a square lattice \cite{11,12,13,14}. 

Additionally, it is
meaningful that consensus models are set up in complex networks.
However, the statistical properties of real-world social networks
vary strongly, for example, the degree distribution can be
single-scale, broad-scale or scale-free\cite{15,16}. Due to the
lack of a single model encompassing the topological features of
social networks, we consider a few established network models
aiming to unveil the effect to different aspects of the topology:
a small-world network (i.e., Watts-Strogatz model \cite{17}) is 
generated by rewiring with a probability the links of a regular 
lattice by long-distance random links\cite{17}; scale-free networks 
(i.e., Barab\'asi-Albert model\cite{18}) are characterized by a 
fat-tailed (power-law) degree distribution and usually modelled by 
growing networks considering a preferential attachment of links; by 
adding a triad formation step on the Barab\'asi-Albert prescription, 
a scale-free model with tunable clustering can be obtained 
(we call it the triad scale-free model)\cite{19}.

Therefore, the aim of our paper is to discuss the original Sznajd
model\cite{11} with synchronous updating on different complex networks:
Watts-Strogatz, Barab\'asi-Albert and triad networks.

\setcounter{footnote}{0}
\renewcommand{\thefootnote}{\alph{footnote}}

\section{The Model}
\noindent On the system lattice, each site $i$ ($i = 1,2, \cdots
,N$, where $N$ is the total number of sites) carries a spin $s_i$, which 
has two possible directions: $s_i=+1$ or $s_i=-1$ . It can be considered 
like an individual that can take one of two possible opinions: 
$s_i  =  + 1$ represents a positive opinion and $s_i  =  - 1 $, a negative 
one. Initially the opinions are distributed randomly, $ + 1$ with probability 
$p$ and $-1$ with probability $1-p$.

The synchronous updating: the system state at time step $t+1$ is
decided by its state at time step $t$. At every time step $t>0$, we 
go systemically through the lattice to find the first member of a pair, 
then the second member of the pair is randomly selected from the neighbors 
of the first one. In this way, one time step means that on average every 
lattice node is selected once as the first member of the pair. The pair 
persuades all its neighbors to assume its (pair) actual opinion at the
next time-step $t+1$, if and only if at the time-step $t$ the pair shares the 
same opinion; otherwise, the neighbor opinions are not affected. In fact, 
after going through the whole lattice once, the time-step $t$ is completed 
and in the beginning of the next time-step ($t+1$), the state of each site 
is updated according to all results of persuasion.

It is important to emphasize that in the synchronous updating, some sites
may feel frustrated and cannot decide the opinion at the next time-step. This 
phenomenon is called as {\bf frustration}. There are two reasons for occurring 
frustration: {\bf (1)} When at the same time-step $t$, an individual is 
persuaded by different pairs to follow different opinions; {\bf (2)} 
If an individual $i$ is selected as member of a pair which persuades the
others to follow its opinion, it intends to keep its opinion unchanged 
at the next time-step $t+1$ ($s_{i,t+1}=s_{i,t}$). However, if at the same
time-step $t$, the individual $i$ is persuaded by other pairs to 
assume an opinion $s$ different from its actual one $s_{i,t}$, then at the 
next time-step $t+1$ the individual $i$ is considered to be an frustrated one.

In our paper, frustrated sites do nothing, i.e., they stay at the time-step 
$t+1$ with the opinion of previous time-step $t$ ($s_{i,t+1}=s_{i,t}$).

On the square lattice, $L \times L$, the Sznajd model with synchronous updating
shows frustration hindering the development of a consensus\cite{11}.
The initial probability $p=0.5$ of opinion $+1$ makes a complete consensus 
much more difficult than $p\ne0.5$. When different initial concentrations 
$p$ of opinion $+1$ are considered, the system shows a transition between the
non-consensus state and the full consensus state \cite{12}. Moreover, there 
is a power-law relation $p \propto L^{ - 0.38} $ between the necessary
value of $p$ to get a full consensus in half the cases and the 
system size $L$. In the following section, we discuss the evolution of 
Sznajd model with synchronous updating in the complex networks.

\section{Simulations of Sznajd model on Watts-Strogatz small-world network}

\begin{figure}[htbp] 
\vspace*{-4pt}
\centerline{\psfig{file=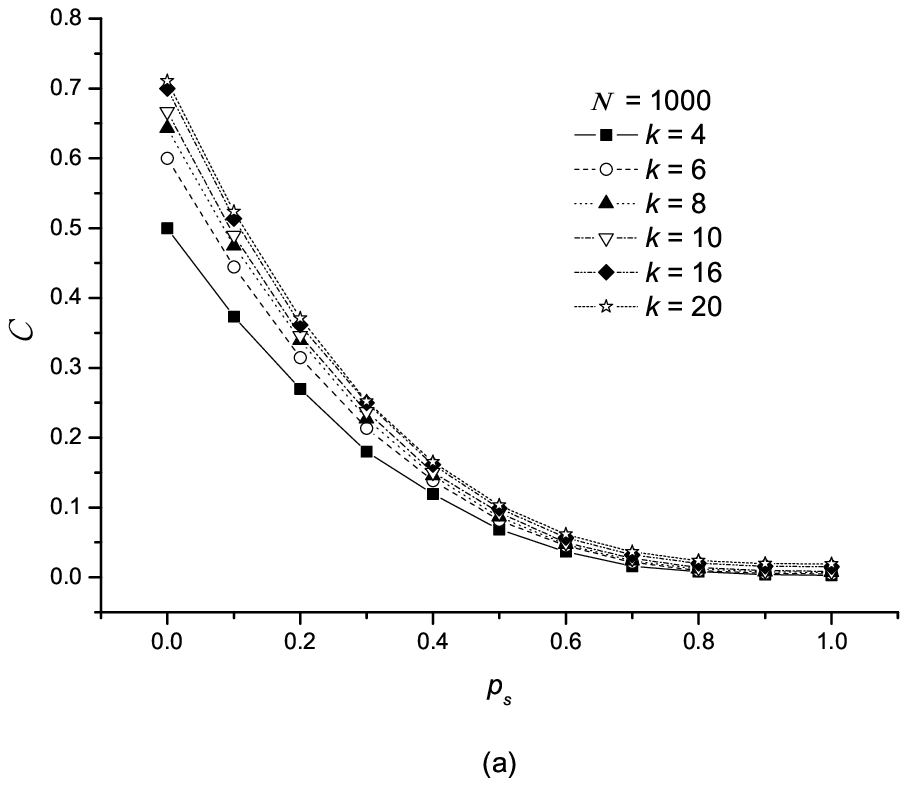,width=6.8cm,height=5.85cm}\hskip
0cm \psfig{file=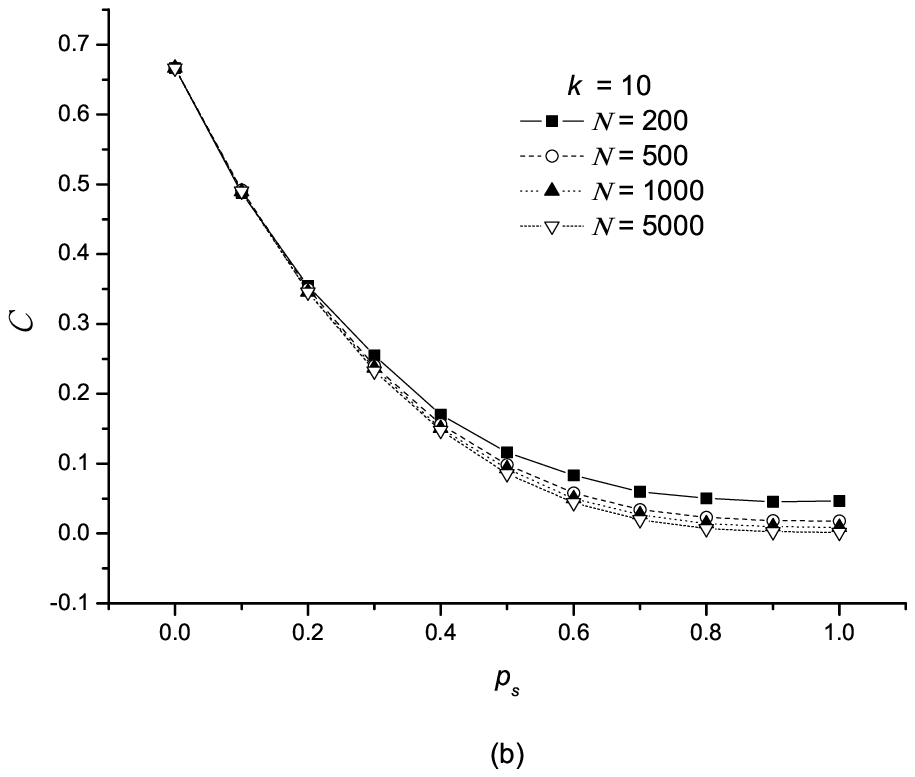,width=6.8cm,height=5.85cm}}
 \vspace*{-3pt} \fcaption{Dependence between $C$ and $p_s $ for the WS
   small-world network.}
\end{figure}
\subsection{Watts-Strogatz small-world network}

\noindent Social networks are far from being regular or completely
random. In the past few years, it has been found that most real-life 
networks have some common characteristics, the most important of which are
called small-world effect and scale-free distribution. The recognition of
small-world effect involves two factors: the clustering coefficient ($C$) 
and the average shortest path length \cite{17,20}; a network is called a
small-world network as long as it has small average shortest path length and
great clustering coefficient. One of the most well-known small-world models 
is Watts-Strogatz small-world network (WS model), which can be constructed 
by the following algorithm: the initial network is a one-dimensional 
lattice of $N$ sites, with periodic boundary conditions (i.e., a ring), 
each site being connected to $k$ nearest neighbors. We choose a vertex and 
the edge that connects it to its nearest neighbor in a clockwise sense. With 
probability $p_s$, we reconnect this edge to a vertex chosen uniformly at
random over the entire ring, with duplicate edges forbidden; otherwise we 
leave the edge in place. This process repeats until one lap is completed and
proceeding outward to more distant neighbors after each lap, until each edge in
the original lattice has been considered once. As it is shown in Figure~1, 
we gain the same dependence between $C \sim p_{s} $  as obtained in 
Ref.~\cite{17}, as well as we can see that when $N>500$ and $k>8$ the 
clustering coefficient ($C$) does not vary very strongly.

\begin{figure}[htbp] 
\vspace*{-4pt}
\centerline{\psfig{file=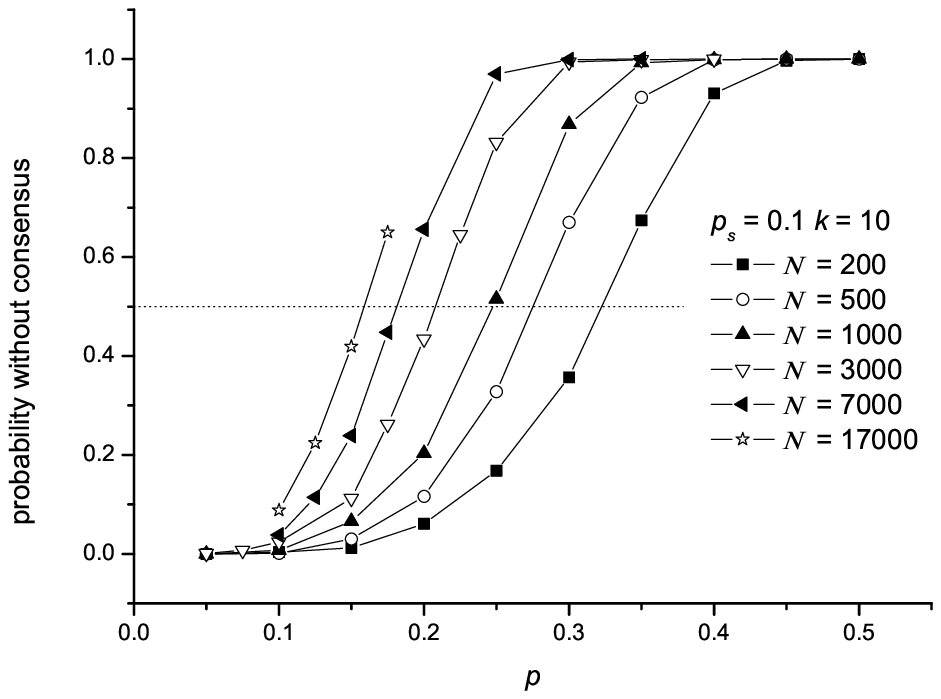,width=7.6cm,height=6.1cm}}
\vspace*{-3pt} \fcaption{Dependence between probability without
consensus and $p$ for the WS small-world network}
\end{figure}

\subsection{The probability of non-consensus as various $p$}
\noindent Every node in the network is considered to be an individual 
with an opinion that in the beginning of the simulation is randomly 
chosen with probability $p$ for opinion $+1$. Once the network has been 
completely constructed, we start the consensus process of Sznajd. With 
up to $1000$ samples ($N \ge 17000$, $400$ samples), similar to the 
square lattice case, frustration hinders the development of a consensus, 
which is not found even after $40000$ time-steps. Figure 2 shows the 
relative number of samples which did not reach a consensus as a function 
of the probability $p$ for initial opinion $+1$. The problem is by
definition symmetric about $p = 0.5$, and only $p \le 0.5$ is thus
plotted in our figures. As it can be seen, we have also observed a 
transition between the non-consensus state and the state with complete 
consensus as function as $p$. When $p = 0.5$, it is most difficult to 
find consensus in the system.

\begin{figure}[htbp] 
\vspace*{-4pt}
\centerline{\psfig{file=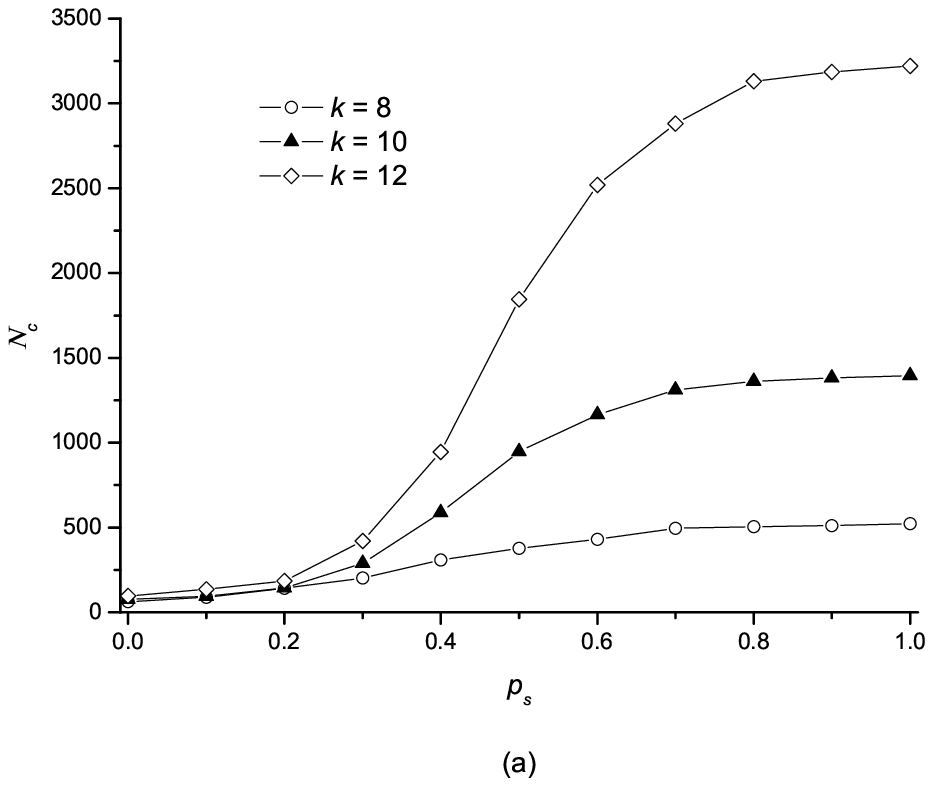,width=6.8cm,height=5.85cm}\hskip
0cm \psfig{file=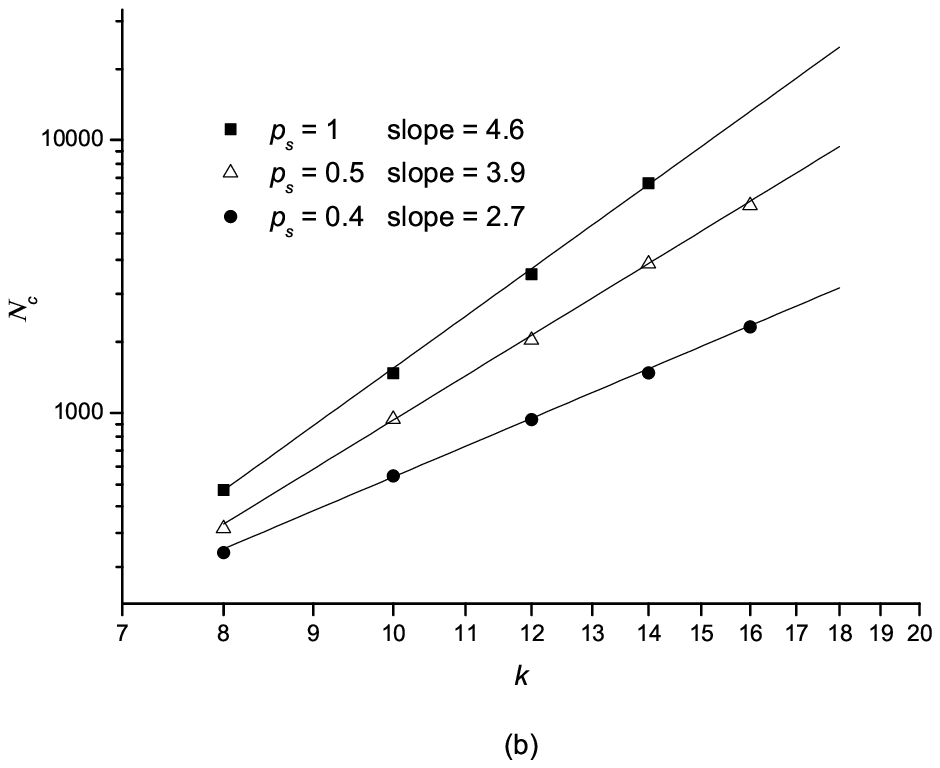,width=6.8cm,height=5.85cm}}
\vspace*{-3pt} \fcaption{Relation between $N_c $ and $p_s$, $k$ for the 
WS small-world network when $p = 0.5$ }
\end{figure}

We have also measured the system size $N_c$ from which we {\bf never} 
find a consensus in any of the samples, i.e, for $N>N_c$ never a full
consensus can be found in any of the samples. Figure 3 shows how this critical 
system size $N_c $ varies with the probability $p_s$ and with the node degree 
$k$. With the rewiring probability $p_s$ increasing at the same $k$, the 
clustering coefficient $C$ decreases (see Figure 1) and $N_c $ increases.
Furthermore, there is a power increase relation between $N_c \sim
k$, and the bigger $p_s $ is, the faster $N_c $ increases (see
Figure 3b). This behavior indicates that a large clustering
coefficient favors the development of a consensus.

\subsection{The necessary value of $p$ to get a consensus in half the cases}
\noindent When $N > N_c $, Figure 4 shows how the $p$ needed to
get a consensus in half the cases varies for different lattice sizes
$N$ and various $p_s$. Since the clustering coefficient $C$ decreases as 
$p_s$ increases (see Figure 1), for $p_{s}<0.5$ and equal-size system $N$, 
$p$ increases and the slope becomes small. However, if $p_{s}>0.5$, the 
tendency of $C$ decreasing slows down (see Figure 1) and according to 
Figure 4, the two lines when $p_s=0.5$ and $p_s=1$ are so close.

\begin{figure}[htbp] 
\vspace*{-4pt}
\centerline{\psfig{file=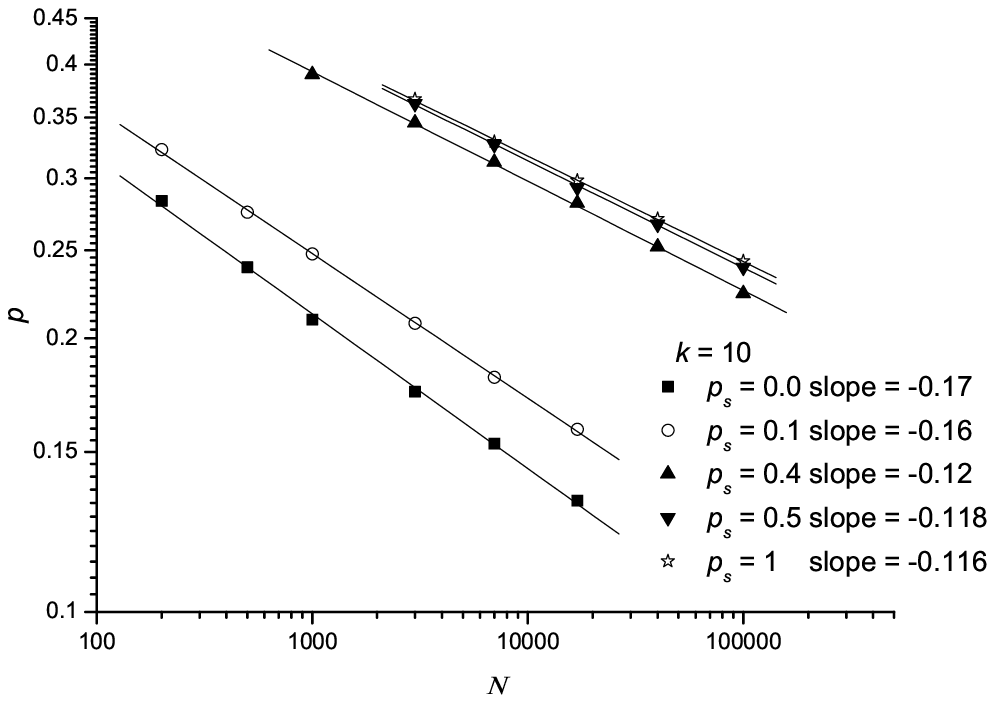,width=7.6cm,height=6.1cm}}
\vspace*{-3pt} \fcaption{Power law relation between the $p$ needed
to get a consensus in half the cases and the system size $N$ for the 
WS small-world network}
\end{figure}

\section{Simulations of Sznajd model on triad scale-free network}
\subsection{Triad scale-free network}
\noindent The small-world networks generated by rewiring links have degree
(number of edges) distributions with exponential tails. In contrast, 
scale-free networks are characterized by a fat-tailed (power-law) degree 
distribution distribution. The most fashionable network presenting both
properties, scale-free and small-world aspects, is the Barab\'asi-Albert
scale-free network (BA network). Although the BA network has successfully 
explained the scale-free nature of many networks, a striking discrepancy 
between it and real networks is that the value of the clustering coefficient 
varies fast with the network size $N$ and for large systems is typically 
several orders of magnitude lower than found empirically (it vanishes in 
the thermodynamic limit \cite{19,21}). In social networks, for instance, 
the clustering coefficient distribution $C(k)$ exhibits a power-law behavior, 
$C(k) \propto k^{-\gamma}$, where $k$ is the node degree (number of neighbors)
and $\gamma \approx 1$ (everyone in the network knows each other).

\begin{figure}[htbp] 
\vspace*{-4pt}
\centerline{\psfig{file=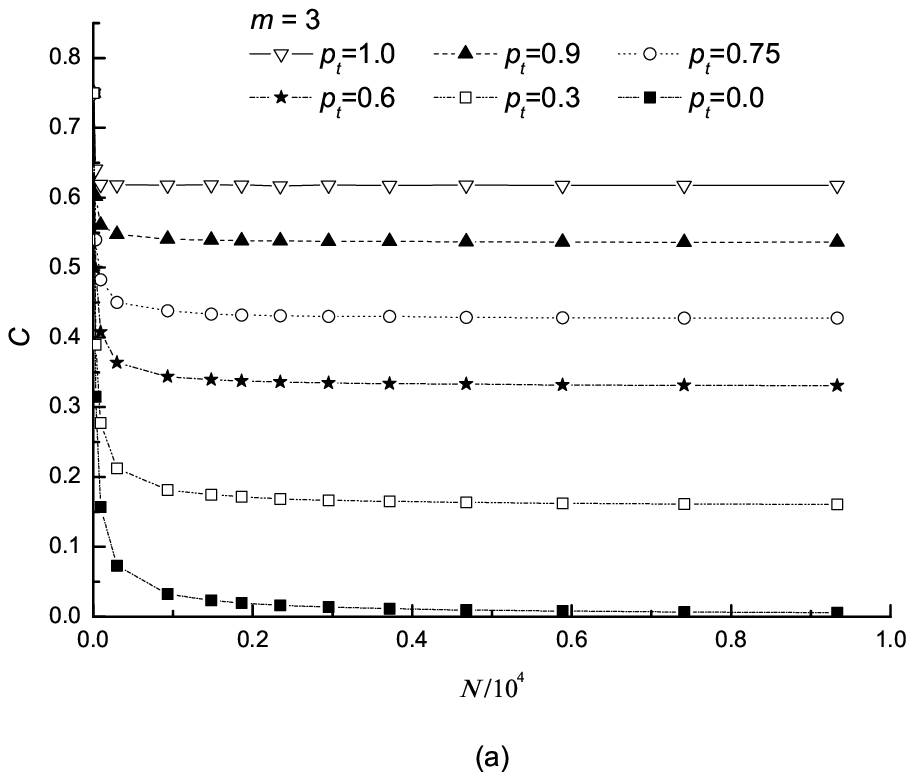,width=6.8cm,height=5.85cm}\hskip
0cm \psfig{file=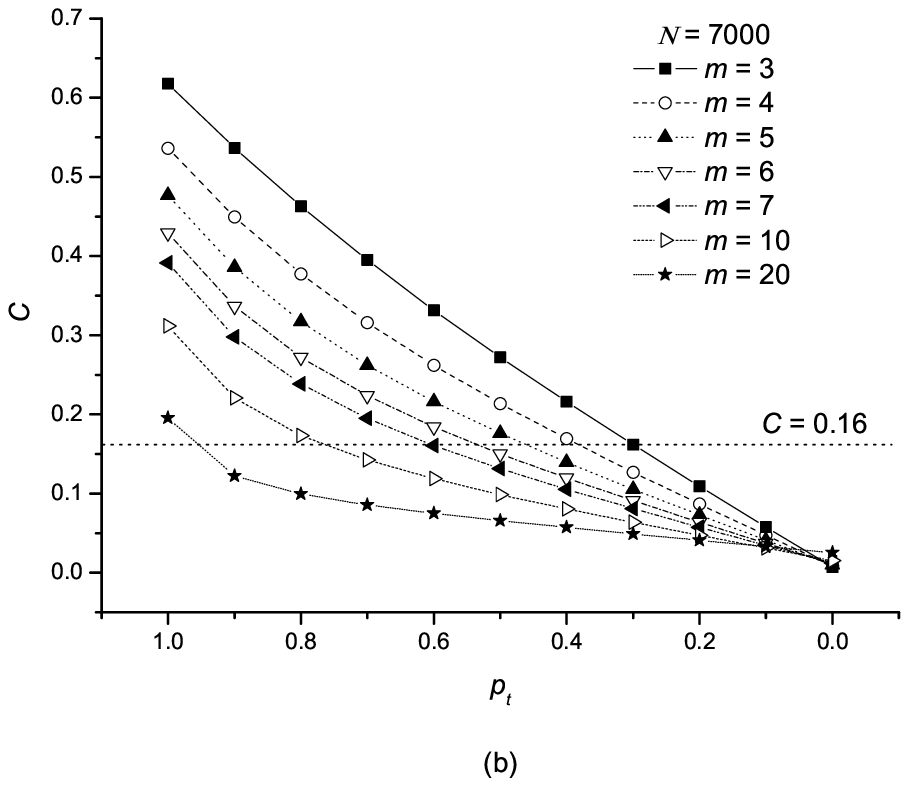,width=6.8cm,height=5.85cm}}
\vspace*{-3pt} \fcaption{Relation between clustering coefficient
$C$ and $N$, $p_t$, $m$ for the triad scale-free network.}
\end{figure}

However, this problem has been surmounted and scale-free models
with high clustering coefficient have been investigated \cite{19},
by adding a triad formation step on the Barab\'asi-Albert
prescription. The Barab\'asi-Albert network starts with a small number 
$m$ of sites all connected with each other. Then a large number $N$ of 
additional sites is added as follows: First, each new node (node $i$) 
performs a preferential attachment step, i.e, it is attached randomly 
to one of the existing nodes (node $j$) with a probability proportional 
to its degree; then follows a triad formation step with a probability 
$p_t$: the new node $i$ selects at random a node in the neighborhood 
of the one linked to in the previous preferential attachment step 
(node $j$). If all neighbors of $j$ are already connected to $i$, 
then a preferential attachment step is performed (``friends of friends 
get friends''). In this model, the original Barab\'asi-Albert network 
corresponds to the case of $p_{t}=0$. It is expected that a nonzero 
$p_t$ gives a finite nonzero clustering coefficient as $N$ is 
increased \cite{19,21}, while the clustering coefficient goes to zero 
when $p_{t}=0$ (the BA scale-free network model), as shown in Figure
5. Indeed, the clustering coefficient increases as the probability $p_t$ and
$m$ increase.

\begin{figure}[htbp] 
\vspace*{-4pt}
\centerline{\psfig{file=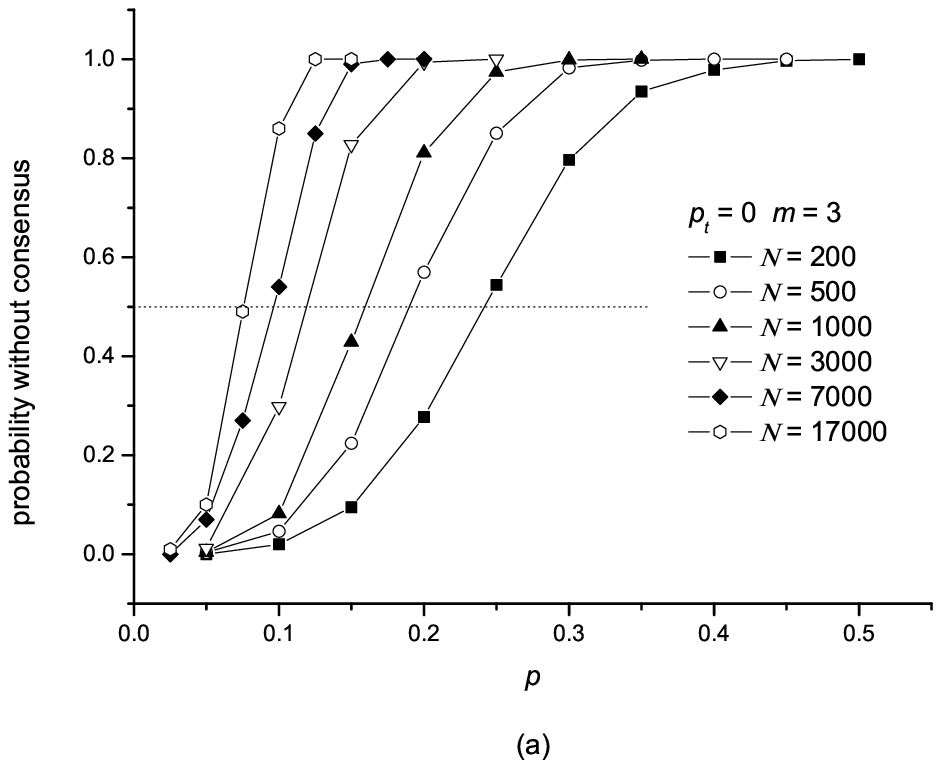,width=6.8cm,height=5.85cm}\hskip
0cm \psfig{file=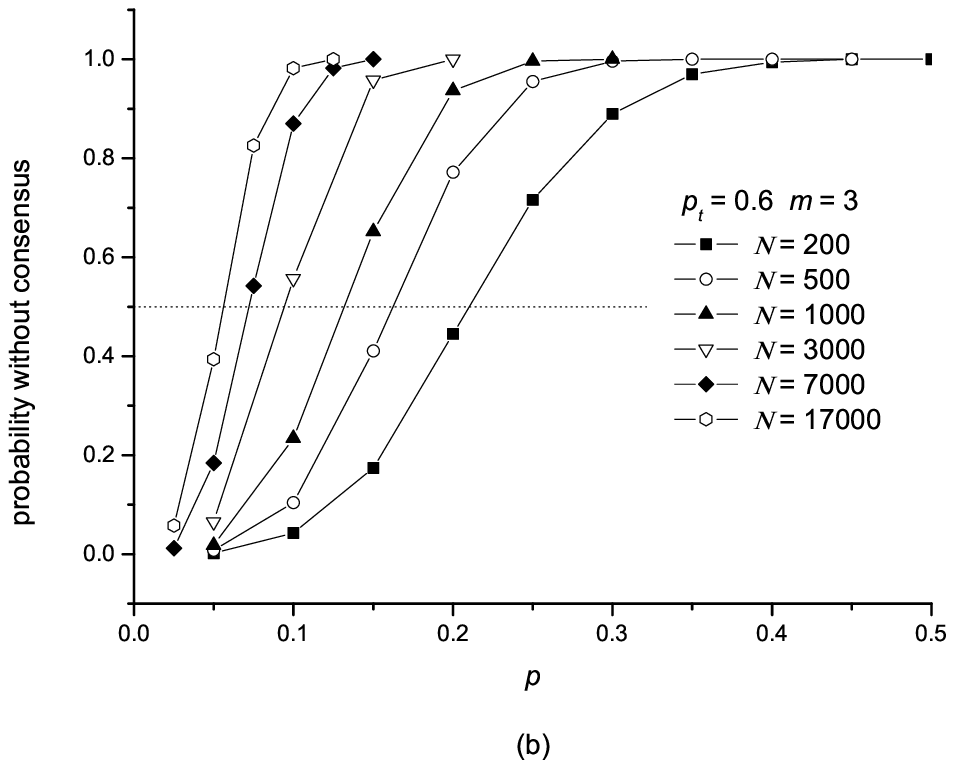,width=6.8cm,height=5.85cm}}
\vspace*{-3pt} \fcaption{Dependence between probability without
consensus and $p$ for the triad scale-free network.}
\end{figure}

\subsection{The probability of non-consensus as various $p$}
\noindent Similar to the WS model, with up to $1000$ samples 
($N \ge 17000$, $400$  samples), a consensus is not found even after 
$40000$ time steps. Figure 6 gives the dependence between the probability 
of non-consensus and $p$. We found a similar transition between the state 
with no-consensus and the state with complete consensus when various $p$ 
for the BA network (Fig. 6a) and for the triad scale-free network (Fig. 6b).

\begin{figure}[htbp] 
\vspace*{-4pt}
\centerline{\psfig{file=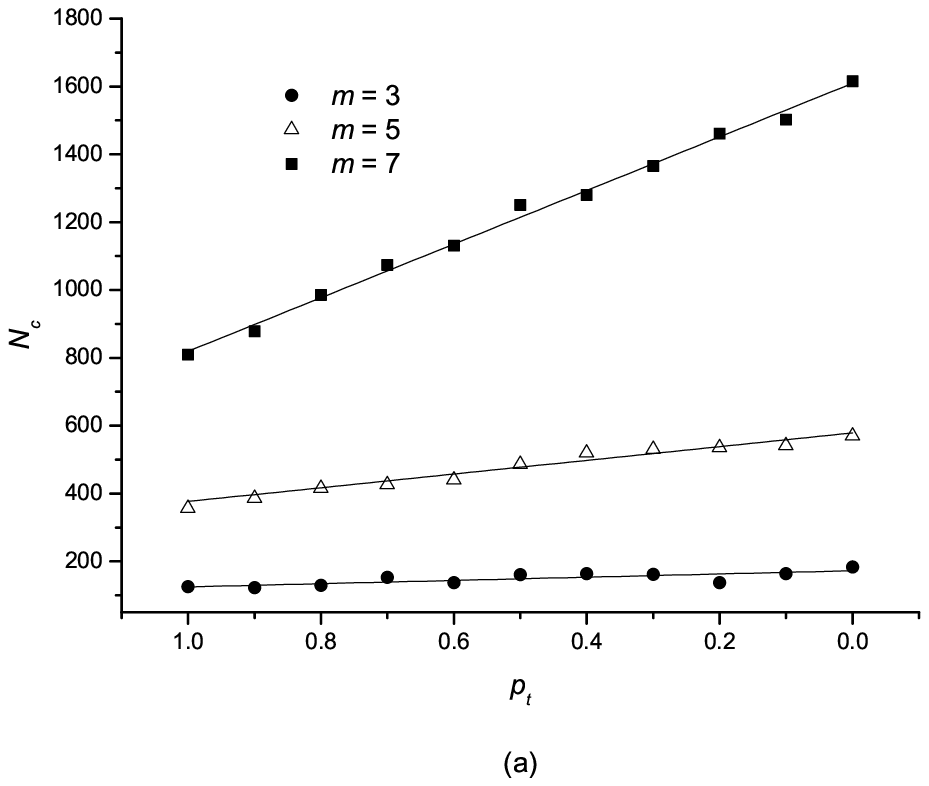,width=6.8cm,height=5.85cm}\hskip
0cm \psfig{file=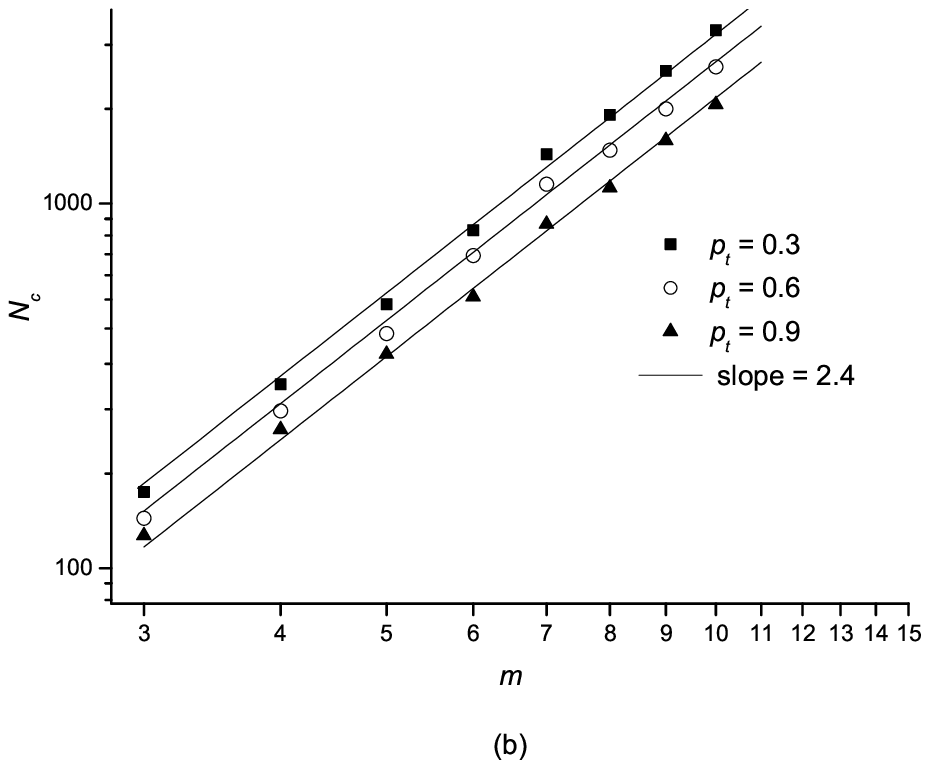,width=6.8cm,height=5.85cm}}
\vspace*{-3pt} \fcaption{Relation between $N_c $ and $ p_t ,m$ for the 
triad scale-free network when $p = 0.5$.}
\end{figure}

In Figure 7, we show the critical system size $N_c$ as a function of 
various values $p_t$, $m$ and $p=0.5$. As $p_t$ increases and $m$ decreases, 
the clustering coefficient $C$ decreases (see Figure 5), thus $N_c$ increases. 
Similar to the WS network, this behavior indicates that a large clustering 
coefficient favors development of a consensus. Different from WS network,
where $N_c \sim p_t$, now $N_c$ follows a power-law $N_{c} \propto m^{2.4}$, 
for different values of $p_t $.

\subsection{The necessary value of $p$ to get a consensus in half the cases}
\noindent When $N > N_c $, Figure 8 shows how the $p$ needed to
get a consensus in half the cases varies for different lattice size
$N$ and various $p_t $. As $p_t$ decreases, the clustering coefficient 
$C$ also decreases (see Figure 5), thus for an equal-size system $N$, $p$ 
increases and the slope becomes smaller (Fig. 8a). Indeed, for a fixed 
value of $p_t$, $p$ decreases as $N$ increases. Since as $m$ increases, 
the clustering coefficient decreases (see Fig. 5b), to compare systems 
with the same clustering coefficient $C=0.16$ and various $m$, the 
probability $p_t$ must be also changed according to the straight line 
in Fig. 5b ($p_t$ increases). For an equal-size system $N$, as $m$ increases 
$p$ increases and the slope becomes bigger (Fig. 8b). As well as for a 
fixed $m$, $p$  decreases as the system size $N$ increases.

\begin{figure}[htbp] 
\vspace*{-4pt}
\centerline{\psfig{file=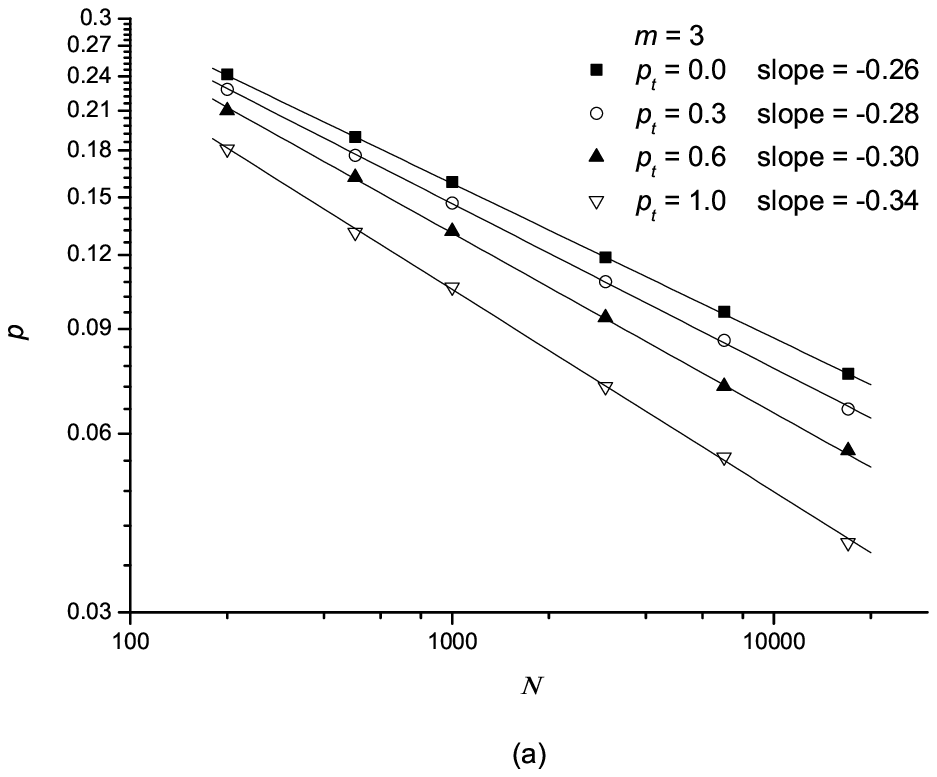,width=6.8cm,height=5.85cm}\hskip
0cm \psfig{file=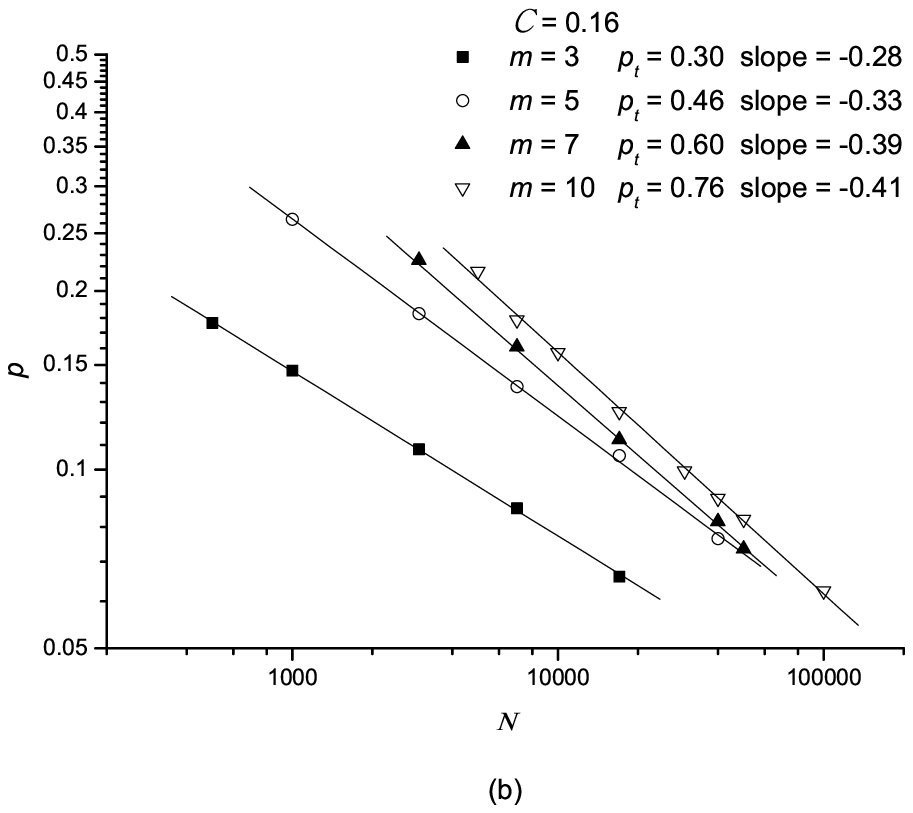,width=6.8cm,height=5.85cm}}
\vspace*{-3pt} \fcaption{Power law relation between the $p$ needed
to get a consensus in half the cases and $N$ for the triad scale-free network.}
\end{figure}

\section{Conclusion}
\noindent Comparing the results of the Sznajd model on a square lattice 
with synchronous updating with our results when the Sznajd model is 
considered on more realistic topologies: Watts-Strogatz small-world network, 
Barab\'asi-Albert scale-free network and triad scale-free network, we notice 
the following similar properties: (1) A transition between the state with
no-consensus and the state with complete consensus; (2) a power
law relation between the initial probability $p$ needed to get a
consensus in half the cases and the system size $N$. However, it
is very interesting to consider the change of these behaviors as
we adjust the parameters of the networks: (1) as $C$ decreases, the 
exponent $\alpha$ of the power-law $p \sim N^{-\alpha}$ decreases; (2) 
as $C$ decreases, the critical system size $N_c $ increases, which 
indicates that a large clustering coefficient favors development of 
a consensus, especially for the triad scale-free network there is a 
power-law relation: $N_c \sim m^{2.4}$. Moreover, in the limit of very 
large networks with $p=0.5$, a consensus seems to be never reached for the WS 
small-world network when we fix $k$ and $p_s$; nor for the scale-free 
network, when we fix $m$ and $p_t$.


\nonumsection{Acknowledgements} \noindent We thank Petter Holme for help, 
and D. Stauffer for a critical reading of the manuscript. This project 
supported by the National Natural Science Foundation of China 
(Grant No.~70371067 and 10347001) and by the Foundation of the Talent 
Project of Guangxi, China (No.~2001204).

\nonumsection{References}

\end{document}